\documentclass[floatfix, preprint, showpacs, showkeys, preprintnumbers, nofootinbib, superscriptaddress]{revtex4-1}
\usepackage[utf8]{inputenc}
\usepackage[sort&compress]{natbib}
\usepackage{ulem}
\usepackage{bm}
\usepackage{times}
\usepackage{amssymb,amsbsy,amsmath,amsfonts}
\usepackage{graphicx}
\usepackage{epstopdf}
\usepackage{float}
\usepackage{color}
\usepackage{morefloats}
\usepackage{rotating}
\usepackage{srcltx}
\usepackage{slashed}
\usepackage{subfigure}
\usepackage{multirow}
\usepackage{verbatim}
\usepackage{hyperref}
\usepackage{overpic}
\usepackage{diagbox}
\usepackage{tabularx}
\usepackage{makecell}

\begin{document}

\title{Hadronic molecules composed of a doubly charmed tetraquark state and a charmed meson  }

\author{Ya-Wen Pan}
\affiliation{School of Physics, Beihang University, Beijing 100191, China}

\author{Tian-Wei Wu}
\affiliation{School of Fundamental Physics and Mathematical Sciences, Hangzhou Institute for Advanced Study, UCAS, Hangzhou 310024, China}
\affiliation{University of Chinese Academy of Sciences, Beijing 100049, China}

\author{Ming-Zhu Liu}\email{zhengmz11@buaa.edu.cn}
\affiliation{School of Space and Environment, Beihang University, Beijing 100191, China}
\affiliation{School of Physics, Beihang University, Beijing 100191, China}

\author{Li-Sheng Geng}\email{lisheng.geng@buaa.edu.cn}
\affiliation{Peng Huanwu Collaborative Center for Research and Education, Beihang University, Beijing 100191, China}
\affiliation{School of Physics, Beihang University, Beijing 100191, China}
\affiliation{
Beijing Key Laboratory of Advanced Nuclear Materials and Physics,
Beihang University, Beijing 100191, China}
\affiliation{School of Physics and Microelectronics, Zhengzhou University, Zhengzhou, Henan 450001, China}

\date{\today}
\begin{abstract}

    The three pentaquark states,  $P_c(4312)$, $P_c(4440)$ and $P_c(4457)$,  discovered by the LHCb Collaboration in 2019, can be arranged into  a complete heavy quark spin symmetry  multiplet of hadronic molecules of $\bar{D}^{(\ast)}\Sigma_{c}^{(\ast)}$. In the heavy quark mass limit, the $\Sigma_{c}^{(\ast)}$ baryons can be related to  the doubly charmed tetraquark states of isospin 1, i.e., $T_{\bar{c}\bar{c}}^{(\ast)}$( $T_{\bar{c}\bar{c}}^{0}$,  $T_{\bar{c}\bar{c}}^{1}$, $T_{\bar{c}\bar{c}}^{2}$),  via heavy antiquark diquark symmetry, which dictates that the $\bar{D}^{(\ast)}\Sigma_{c}^{(\ast)}$ interactions  are the same as  the $\bar{D}^{(\ast)}T_{\bar{c}\bar{c}}^{(\ast)}$ interactions up to { heavy antiquark diquark symmetry}  breakings.      In this work,  we   employ the contact-range effective field theory to systematically study the  $\bar{D}^{(\ast)}T_{\bar{c}\bar{c}}^{(\ast)}$ systems, and we show the existence of a complete heavy quark spin symmetry multiplet of hadronic molecules composed of a
    doubly charmed tetraquark state and a charmed meson.    These are a new kind of hadronic  molecules and, if discovered, can lead to a better understanding of the many exotic hadrons discovered so far.  In addition, {   we summarise the triply charmed hexaquark states formed by different combinations of hadrons. In particular, we show that $\bar{\Omega}_{ccc}{p}$ system can bind by the Coulomb force, which is analogous  to a hydrogenlike atom. }

\end{abstract}

%\pacs{13.60.Le, 12.39.Mk,13.25.Jx}

\maketitle

\section{Introduction}

A hadronic molecule is a bound  or  resonant state composed of conventional hadrons that is held together by the residual strong interaction, which is analogous to the deuteron that is bound by the nuclear force. The studies on hadronic molecules containing charm quarks began with the proposal that $\psi(4040)$ is  a $\bar{D}^{\ast}D^{\ast}$ hadronic molecule~\cite{DeRujula:1976zlg,Voloshin:1976ap}. Although nowadays $\psi(4040)$ is widely viewed as a conventional charmonium state, the idea motivated a lot of theoretical studies on heavy hadronic molecules~\cite{Tornqvist:1991ks,Tornqvist:1993ng,Manohar:1992nd,Ericson:1993wy}. Such studies were revived   after the discovery of $X(3872)$ by the Belle Collaboration in 2003~\cite{Belle:2003nnu}.  Treated as as a $D\bar{D}^{\ast}$ bound state, the lower mass (than the conventional quark model prediction) and the isospin  violation of the strong decays of $X(3872)$ can be easily understood, which implies that $X(3872)$ contains a large molecular component~\cite{Gamermann:2009fv,Li:2012cs,Wu:2021udi}. Since then a series of states beyond the naive quark model picture have been discovered and many of them can be regarded as  molecular candidates. See Refs.~\cite{Chen:2016qju,Hosaka:2016ypm,Lebed:2016hpi,Guo:2017jvc,Olsen:2017bmm,Ali:2017jda,Brambilla:2019esw,Liu:2019zoy,Chen:2022asf} for some recent reviews.   

Symmetry is a particularly important and useful concept in particle physics. The SU(3)-flavor symmetry has already achieved great successes, e.g.,  in classifying the ground-state  mesons and baryons. Symmetry plays an important role in studies of hadronic molecules as well, such as SU(3)-flavor and  heavy quark spin symmetry (HQSS).   Assuming that  $D_{s0}(2317)$ discovered by the BarBar  Collaboration~\cite{BaBar:2003oey}  is a $DK$ molecule,   it is natural to expect a  $D^{\ast}K$ molecule in terms of HQSS, corresponding to the exotic state $D_{s1}(2460)$~\cite{Altenbuchinger:2013vwa,Guo:2006rp,Bali:2017pdv,Hu:2020mxp}. 
The $Z_{c}(3900)$ and $Z_{cs}(3985)$ states,  discovered by the BESIII Collaboration~\cite{BESIII:2013ris,BESIII:2020qkh}, are  often interpreted as $\bar{D}D^{\ast}$ and  $\bar{D}D_{s}^{\ast}/\bar{D}_{s}D^{\ast}$ molecules, and  are related to each other by  SU(3)-flavor symmetry~\cite{Meng:2020ihj,Yang:2020nrt,Wang:2020htx,Yan:2021tcp,Du:2022jjv,Baru:2021ddn,Zhai:2022ied}.  Another example is the
three hidden charm pentaquark states, $P_{c}(4312)$,   $P_{c}(4440)$ and $P_{c}(4457)$, discovered by the LHCb Collaboration~\cite{Aaij:2019vzc}. They  can be  arranged into  a complete HQSS multiplet  of $\bar{D}^{(\ast)}\Sigma_{c}^{(\ast)}$ hadronic molecules~\cite{Xiao:2019aya,Sakai:2019qph,Yamaguchi:2019seo,Liu:2019zvb,Valderrama:2019chc,Meng:2019ilv,Du:2019pij}. The existence of such multiplets of hadronic molecules not only enriches the hadron spectrum, but also can help verify the molecular nature of some exotic states~\cite{Liu:2020hcv,Liu:2020tqy}. 

The heavy antiquark diquark symmetry(HADS) dictates that  a pair of heavy quarks behaves like a heavy antiquark from the perspective of the color degree of freedom~\cite{Savage:1990di}. Therefore,  the charmed antimesons are  the same  as the  doubly charmed baryons. Via HADS, one can derive the following relation between the mass splitting of the doubly charmed baryon doublet   and that of the charmed meson doublet, i.e., $m_{\Xi_{cc}^*}-m_{\Xi_{cc}}=\frac{3}{4}(m_{\bar{D}^*}-m_{\bar{D}})$~\cite{Hu:2005gf}, which has been verified by a series of lattice QCD simulations~\cite{Padmanath:2015jea,Chen:2017kxr,Alexandrou:2017xwd,Mathur:2018rwu}.
In addition, the coupling of a doubly charmed baryon  to a pion can be related to that of a charmed meson  to a pion via HADS, which is found to be consistent with the quark model prediction~\cite{Liu:2018bkx,Liu:2018euh}. Applying HADS to the $\Sigma_{c}^{(\ast)}$ baryons, it is natural to expect the existence of a set of compact doubly charmed tetraquark states~\cite{Eichten:2017ffp}, which have received renewed interests since the recent discovery of  the doubly charmed tetraquark  $T_{cc}$ by the LHCb Collaboration~\cite{LHCb:2021vvq,LHCb:2021auc}. 
In Ref.~\cite{Cheng:2020wxa},  HADS has been utilized to predict the existence of a series of compact doubly charmed tetraquark states as well. 

\begin{figure}[!h]
\begin{center}
\includegraphics[width=5.3in]{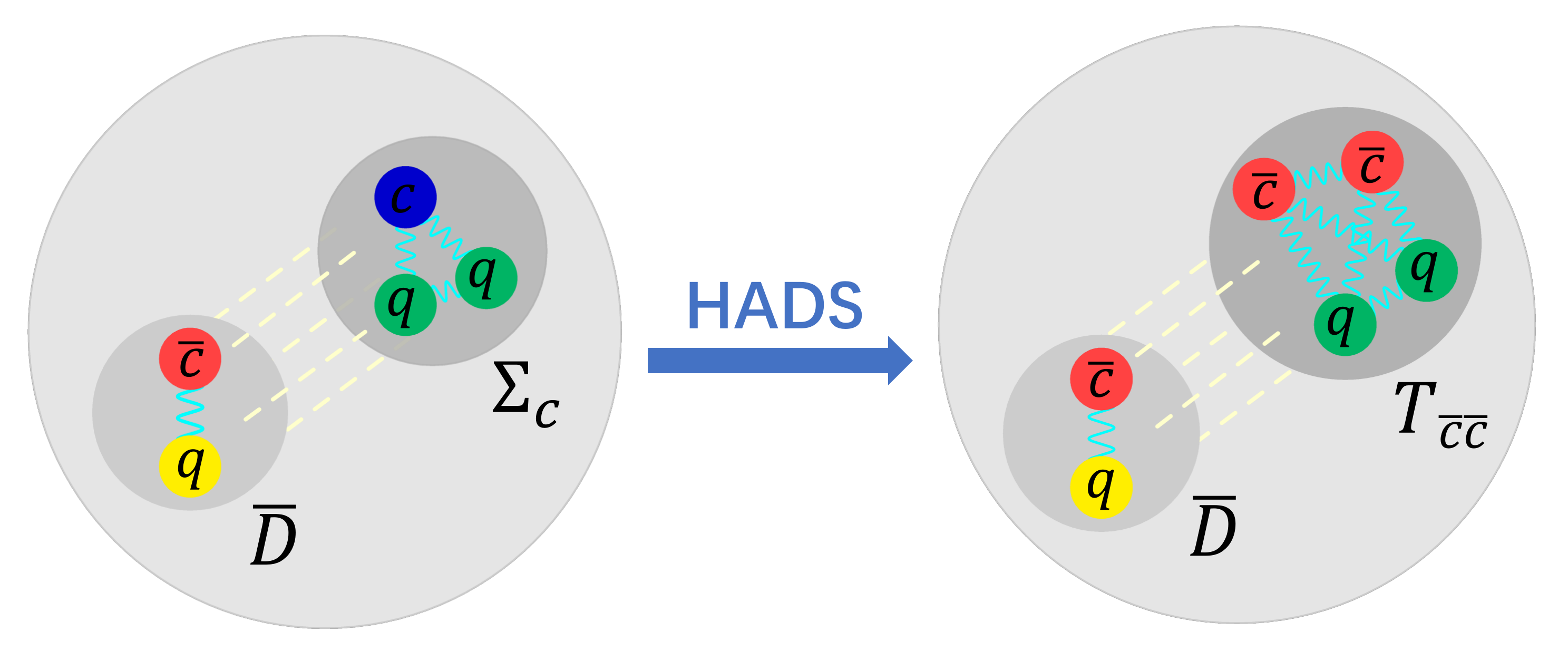}
\caption{From the $\bar{D}^{(\ast)}\Sigma_{c}^{(\ast)}$ system to the $\bar{D}^{(\ast)}T_{\bar{c}\bar{c}}^{(\ast)}$ system via HADS.   }
\label{DN}
\end{center}
\end{figure}

The HADS has also been  utilized  to study hadronic molecules.  In Refs.~\cite{Guo:2013xga,Liu:2020tqy},   the authors related     the $D^{(\ast)}\Xi_{cc}^{(\ast)}$ and $\bar{\Xi}_{cc}^{(\ast)}\Xi_{cc}^{(\ast)}$ systems with the $\bar{D}^{(\ast)}D^{(\ast)}$ system  via HADS, and investigated the likely existence of hadronic molecules of $D^{(\ast)}\Xi_{cc}^{(\ast)}$ and $\bar{\Xi}_{cc}^{(\ast)}\Xi_{cc}^{(\ast)}$.
Given that there exists a complete multiplet of HQSS hadronic molecules of  $\bar{D}^{(\ast)}\Sigma_{c}^{(\ast)}$,   it is natural to expect the existence of    $\Xi_{cc}^{(\ast)}\Sigma_{c}^{(\ast)}$ hadronic molecules by replacing  $\bar{D}^{(\ast)}$ with $\Xi_{cc}^{(\ast)}$ via HADS~\cite{Liu:2018zzu,Pan:2019skd}. In this work, following Ref.~\cite{Pan:2019skd}, we relate the $\bar{D}^{(\ast)}T_{\bar{c}\bar{c}}^{(\ast)}$ system to the  $\bar{D}^{(\ast)}\Sigma_{c}^{(\ast)}$ system via HADS as shown in Fig.~\ref{DN}, and then employ the contact-range effective field theory (EFT) to investigate the  likely existence of $\bar{D}^{(\ast)}T_{\bar{c}\bar{c}}^{(\ast)}$ hadronic molecules. Such molecules composed of a charmed meson and a compact doubly charmed tetraquark state are a new kind of  triply charmed hexaquark states in  addition to the dibaryon  states  $\Xi_{cc}^{(\ast)}\Sigma_{c}^{(\ast)}$~\cite{Pan:2019skd} and three-body molecules $D^{(\ast)}D^{(\ast)}D^{(\ast)}$~\cite{Wu:2021kbu,Luo:2021ggs}. 

This paper is organized as follows.  In Sec.~\ref{sec:formalism}, we explain how to derive the contact potential for the $\bar{D}^{(\ast)}T_{\bar{c}\bar{c}}^{(\ast)}$ system constrained  by HQSS, solve the  Lippmann-Schwinger equation, and search for poles which correspond to $\bar{D}^{(\ast)}T_{\bar{c}\bar{c}}^{(\ast)}$ molecules. In Sec.~\ref{sec:pre}, we present the pole positions  obtained with both  single-channel and coupled-channel potentials.   Finally, this paper is ended with a short summary in Sec.~\ref{sum}.

\section{Theoretical formalism}
\label{sec:formalism}

In this section, we explain how to derive the contact-range potential for the  $\bar{D}^{(\ast)}T_{\bar{c}\bar{c}}^{(\ast)}$ system constrained by HQSS.    {  Here we note that in terms of HADS  the isospin of the $\bar{D}^{(\ast)}T_{\bar{c}\bar{c}}^{(\ast)}$ system is $1/2$ because the isospin of the $\bar{D}^{(\ast)}\Sigma_{c}^{(\ast)}$ system is $1/2$.      }   The $\Sigma_{c}^{(\ast)}$ baryons are made of a charm quark and a pair of light quarks in the conventional quark model.  HADS dictates that the charm quark of a $\Sigma_{c}^{(\ast)}$ baryon can be replaced with a pair of anti-charm quarks, and then it transforms  to a doubly charmed tetraquark state as shown in Fig.~\ref{DN}, where  the pair of anti-charm quarks is in the color  $\bar{3}$ state of spin 1. Combining with the spin and color configurations of a pair of light quarks, the  doubly charmed tetraquark states can have spin 0, 1, or 2, i.e.,
\begin{eqnarray}
|T_{\bar{c}\bar{c}}(0^{+})\rangle &=& 1_{H}^{3}\otimes
{1}_{L}^{\bar{3}},     \\ \nonumber
|T_{\bar{c}\bar{c}}(1^{+})\rangle &=&
1_{H}^{3}\otimes
{1}_{L}^{\bar{3}},  \\ \nonumber
|T_{\bar{c}\bar{c}}(2^{+})\rangle &=&
1_{H}^{3}\otimes
{1}_{L}^{\bar{3}},  
\end{eqnarray}
where $1_{H}^{3}$ and ${\color{red}1_{L}^{\bar{3}}}$ represent an anti-charm quark pair and a light quark pair, respectively.   
In the following, we denote $T_{\bar{c}\bar{c}}(0^{+})$, $T_{\bar{c}\bar{c}}(1^{+})$, and $T_{\bar{c}\bar{c}}(2^{+})$ as $T_{\bar{c}\bar{c}}^{0}$, $T_{\bar{c}\bar{c}}^{1}$, and $T_{\bar{c}\bar{c}}^{2}$, respectively. {  The  existence of isovector tetraquark states $T_{\bar{c}\bar{c}}^{(\ast)}$ not only is allowed by Quantum Chromodynamics (QCD) but also were predicted by a series of theoretical studies~\cite{Ader:1981db,Vijande:2003ki,Ebert:2007rn,Luo:2017eub,Eichten:2017ffp,Karliner:2017qjm}. In this work,   we 
 take the HADS  to argue for their existence. The recent theoretical studies  have shown that the masses of such isovector $T_{\bar{c}\bar{c}}^{(\ast)}$ states are above the mass thresholds of a pair of relevant charmed mesons~\cite{Lu:2020rog,Cheng:2020wxa,Guo:2021yws,Weng:2021hje,Kim:2022mpa}, which implies that such states would decay into $DD\pi$ via an on-shell $D^*$ meson,  leading  to  larger widths than that of  $T_{cc}^{+}(3875)$~\cite{DelFabbro:2004ta}.
 As a result, it will be necessary for the LHCb Collaboration to  update their data sample and enlarge the energy region studied  to search for  these isovector $T_{\bar{c}\bar{c}}^{(\ast)}$ states.   } Since these states have not been discovered experimentally, we turn to the theoretical works~\cite{Cheng:2020wxa,Kim:2022mpa,Zhang:2021yul,Weng:2021hje} for their masses, which are given in Table~\ref{masses}. In the following numerical study, we use their averages given in the last column of Table~\ref{masses}.

\begin{table}[!h]
\centering
\caption{Masses (in units of MeV) of  $T_{\bar{c}\bar{c}}^{1}$, $T_{\bar{c}\bar{c}}^{2}$, and $T_{\bar{c}\bar{c}}^{3}$ in several models and their averages. \label{masses}
}
\label{results}
\begin{tabular}{c c c c c c c c}
  \hline \hline
    Tetraquark    &~~~~\cite{Cheng:2020wxa}  &~~~\cite{Kim:2022mpa}   &~~~\cite{Zhang:2021yul}  &~~~\cite{Weng:2021hje}   &~~~ A.V
         \\ \hline 
       $T_{\bar{c}\bar{c}}^{0}$   &~~~~ 3999.8&~~~4132 &~~~4032  &~~~3969.2  &~~~4033.3
         \\   $T_{\bar{c}\bar{c}}^{1}$   &~~~~ 4124.0&~~~4151 &~~~~4117 &~~~4053.2 &~~~4111.3
         \\  $T_{\bar{c}\bar{c}}^{2}$ &~~~~ 4194.9&~~~4185 &~~~~4179&~~~4123.8&~~~4170.7 
         \\   
  \hline \hline
\end{tabular}
\end{table}

The HQSS dictates  that the interactions between light quarks are independent of the spin of the heavy quark(s).
In the following, we apply HQSS to derive the $\bar{D}^{(\ast)}T_{\bar{c}\bar{c}}^{(\ast)}$ contact-range potentials. First, we express  the spin wave functions of the $\bar{D}^{(\ast)}T_{\bar{c}\bar{c}}^{(\ast)}$ pairs  in terms of the spins of the heavy quarks $s_{1H}$ and $s_{2H}$ and those of the light quark(s) (often referred to as brown muck~\cite{Isgur:1991xa,Flynn:1992fm}) $s_{1L}$ and $s_{2L}$, where 1 and 2 denote $\bar{D}^{(*)}$ and $T_{\bar{c}\bar{c}}^{(*)}$, respectively, via the following spin coupling formula,
\begin{eqnarray}
&&|s_{1l}, s_{1h}, j_{1}; s_{2l}, s_{2h},j_{2}; J\rangle =
  \\ \nonumber &&
  \sqrt{(2j_{1}+1)(2j_{2}+1)(2s_{L}+1)(2s_{H}+1)}\left(\begin{matrix}
s_{1l} & s_{2l} & s_{L} \\
s_{1h} & s_{2h} & s_{H} \\
j_{1} & j_{2} & J%
\end{matrix}\right)|s_{1l},
s_{2l}, s_{L}; s_{1h}, s_{2h},s_{H}; J\rangle,
\label{9j}
\end{eqnarray}
where $j_{1}$ and $j_{2}$ represent the spin of $\bar{D}^{(*)}$ and $T_{\bar{c}\bar{c}}^{(\ast)}$, $S_{L}$ and $S_{H}$ represent the total spin of the light quarks and heavy quarks, and $J$ represent the total angular momentum of the $\bar{D}^{(\ast)}T_{\bar{c}\bar{c}}^{(\ast)}$ system. Ten  $\bar{D}^{(\ast)}T_{\bar{c}\bar{c}}^{(\ast)}$ states can be decomposed into the light quark basis $S_{L}$ and heavy quark basis  $S_{H}$, which are explicitly given   in Appendix A.

In the heavy quark mass limit,  the $\bar{D}^{(\ast)}T_{\bar{c}\bar{c}}^{(\ast)}$ potentials can be parameterized by two coupling constants  describing the interactions between light quarks of spin 1/2 and 3/2, respectively, i.e.,  $F_{1/2}=\langle 1/2_{L} | V| 1/2_{L} \rangle$  and $F_{3/2}=\langle 3/2_{L} | V| 3/2_{L} \rangle$, which are explicitly  given in Appendix A.  The $\bar{D}^{(\ast)}T_{\bar{c}\bar{c}}^{(\ast)}$ potentials can be rewritten as a combination of $C_{a}$ and $C_{b}$, i.e. $F_{1/2} = C_a-2C_b$ and  $F_{3/2} = C_a+C_b$~\cite{Liu:2019tjn}.
For the $T_{\bar{c}\bar{c}}^{0}\bar{D}-T_{\bar{c}\bar{c}}^{1}\bar{D}^{\ast}$ coupled channels, the contact-range potentials $V$ in matrix form read 
\begin{equation}
    V_{T_{\bar{c}\bar{c}}^{0}\bar{D}-T_{\bar{c}\bar{c}}^{1}\bar{D}^{\ast}}^{J=0}=\begin{pmatrix}C_a&-\sqrt{2}C_b
    \\-\sqrt{2}C_b&C_a -C_b \end{pmatrix}.
\end{equation}
For the $T_{\bar{c}\bar{c}}^{1}\bar{D}-T_{\bar{c}\bar{c}}^{0}\bar{D}^{\ast}-T_{\bar{c}\bar{c}}^{1}\bar{D}^{\ast}-T_{\bar{c}\bar{c}}^{2}\bar{D}^{\ast}$ coupled channels, the contact-range potentials $V$ in matrix form read 
\begin{equation}
    V_{T_{\bar{c}\bar{c}}^{1}\bar{D}-T_{\bar{c}\bar{c}}^{0}\bar{D}^{\ast}-T_{\bar{c}\bar{c}}^{1}\bar{D}^{\ast}-T_{\bar{c}\bar{c}}^{2}\bar{D}^{\ast}}^{J=1}=\begin{pmatrix}
       C_a&\sqrt{\frac{2}{3}}C_b  & -\frac{1}{\sqrt{2}}C_b  & -\sqrt{\frac{5}{6}}C_b
    \\\sqrt{\frac{2}{3}}C_b &C_a  &\frac{2}{ \sqrt{3}}C_b  & 0
    \\-\frac{1}{\sqrt{2}}C_b&\frac{2}{ \sqrt{3}}C_b  & C_a-\frac{1}{2}C_b & \frac{1}{2}\sqrt{\frac{5}{3}}C_b  \\-\sqrt{\frac{5}{6}}C_b&0  & \frac{1}{2}\sqrt{\frac{5}{3}}C_b & C_a-\frac{3}{2}C_b
    \end{pmatrix}.
\end{equation}
For the $T_{\bar{c}\bar{c}}^{2}\bar{D}-T_{\bar{c}\bar{c}}^{1}\bar{D}^{\ast}-T_{\bar{c}\bar{c}}^{2}\bar{D}^{\ast}$ coupled channels, the contact-range potentials $V$ in matrix form read 
\begin{equation}
    V_{T_{\bar{c}\bar{c}}^{2}\bar{D}-T_{\bar{c}\bar{c}}^{1}\bar{D}^{\ast}-T_{\bar{c}\bar{c}}^{2}\bar{D}^{\ast}}^{J=2}=\begin{pmatrix}
C_a  &\frac{1}{ \sqrt{2}}C_b  & -\frac{3}{\sqrt{6}}C_b
    \\\frac{1}{ \sqrt{2}}C_b  & C_a+\frac{1}{2}C_b & \frac{\sqrt{3}}{2}C_b  \\-\frac{3}{\sqrt{6}}C_b & \frac{\sqrt{3}}{2}C_b  & C_a-\frac{1}{2}C_b
    \end{pmatrix}.
\end{equation}
For the $T_{\bar{c}\bar{c}}^{2}\bar{D}^{\ast}$ single channel, the contact-range potential is $V_{T_{\bar{c}\bar{c}}^{2}\bar{D}^{\ast}}^{J=3}=C_a+C_b$. {  In the present work, we denote the diagonal and off-diagonal  elements  of  the above matrices    as elastic potentials and inelastic potentials of the $\bar{D}^{(\ast)}T_{\bar{c}\bar{c}}^{(\ast)}$  system, respectively.    } 

With the above potentials, we can solve the   Lippmann-Schwinger equation
\begin{eqnarray}
T=(1-VG)^{-1}V,
\end{eqnarray}
where $V$ is the coupled-channel potential determined by the contact-range EFT approach described above, and $G$ is the two-body propagator. 
In evaluating the loop function $G$, we introduce a regulator of Gaussian form $e^{-2q^{2}/\Lambda^2}$ in the integral as
\begin{eqnarray}
G(s)=\int \frac{d^{3}q}{(2\pi)^{3}} \frac{e^{-2q^{2}/\Lambda^2}}{{\sqrt{s}}-m_{1}-m_{2}-q^{2}/(2\mu_{12})+i \varepsilon}
\label{loopfunction},
\end{eqnarray}
where { $\sqrt{s}$ }is  the total energy in the center-of-mass (c.m.) frame of $m_{1}$ and $m_{2}$, $\mu_{12}=\frac{m_{1}m_{2}}{m_{1}+m_{2}}$ is the reduced mass, and $\Lambda$ is the momentum cutoff. Following our previous works~\cite{Liu:2019tjn,Xie:2022hhv}, we take $\Lambda=0.75$~GeV and $\Lambda=1.5$~GeV in the present work, which reflects the internal structure of the involved hadrons and provides an estimate of the uncertainties of the numerical results.        The dynamically generated states correspond to poles in  the unphysical  sheet.  In this sheet, the loop function of Eq.~(\ref{loopfunction}) becomes
\begin{eqnarray}
G^{II}(s,m_1,m_2)=G(s,m_1,m_2)+i\mu_{12} \frac{p  }{2\pi}{  e^{-2 p^2/\Lambda^2}},
\end{eqnarray}
where the c.m. momentum $p$  is  
\begin{eqnarray}
p=\sqrt{2\mu_{12}\left(\sqrt{s}-m_1-m_2\right)}.
\end{eqnarray}

\section{Results and Discussions}
\label{sec:pre}

Following our previous works~\cite{Liu:2019tjn,Pan:2019skd}, 
the parameters $C_a$ and $C_b$ of the contact-range potentials  are  determined by reproducing the masses of $P_{c}(4440)$ and $P_{c}(4457)$ in two scenarios.  In Scenario A,  $P_{c}(4440)$ and $P_{c}(4457)$ have  $J^P=1/2^{-}$ and $J^P=3/2^{-}$, while   in Scenario B the assignment interchanges. In Table~\ref{couplingconstants},  we present the values of $C_a$ and $C_b$ obtained in Scenario A and B with  $\Lambda=1.5$ GeV and $\Lambda=0.75$ GeV . One should note that the uncertainty of HADS is expected to be of the order of $\Lambda_{QCD}/(m_{Q}v)$~\cite{Savage:1990di}, where $v$ is the velocity of the heavy quark pair. In general, the QCD scale is taken to be $\Lambda_{QCD}\sim200-300$ MeV and $m_{Q}v$ is estimated to be $800$ MeV for charm quarks.  The breaking of HADS is then estimated to be
$25-40\%$~\cite{Pan:2019skd}. The mass splitting between the $J=1/2$ and $J=3/2$ doubly charmed baryons obtained in lattice QCD simulations are $20-25\%$ smaller than the HADS prediction~\cite{Padmanath:2015jea,Chen:2017kxr,Alexandrou:2017xwd,Mathur:2018rwu}. Therefore, we settle on a $25\%$ uncertainty for the $\bar{D}^{(\ast)}T_{\bar{c}\bar{c}}^{(\ast)}$  potentials.

\begin{table}[ttt]
\centering
\caption{ Couplings $C_a$ and $C_b$ (in units of GeV$^{-2}$) in Scenario A and B obtained with  $\Lambda=0.75$ GeV and $\Lambda=1.50$ GeV.
}
\label{couplingconstants}
\begin{tabular}{c| c c c| c c c }
  \hline \hline
     Scenario   &~~~~ $\Lambda$ (GeV)   &~~~~~~ $C_a$  &~~~~   $C_b$  &~~~~ $\Lambda$ (GeV)   &~~~~ $C_a$  &~~~~   $C_b$ 
         \\ \hline  A   &~~~~ $1.50$  &~~~~ $-12.18$   &~~~~~~$1.07$  &~~~~ $0.75$   &~~~~ $-30.80$   &~~~~   $5.10$ 
         \\   B   &~~~~ $1.50$  &~~~~ $-12.89 $  &~~~~  $-1.07$   &~~~~$0.75$  &~~~~ $-34.20$   &~~~~   $-5.10 $ 
         \\
  \hline \hline
\end{tabular}
\end{table}

\begin{table}[ttt]
\centering
\caption{Binding energies (in units of MeV) of the $\bar{D}^{(\ast)}T_{\bar{c}\bar{c}}^{(\ast)}$ molecules  in Scenario A and Scenario B obtained with single-channel potentials. The numbers inside and outside the brackets correspond to $\Lambda=0.75$ GeV and $\Lambda=1.5$ GeV. The superscripts and subscripts are obtained by allowing for a $25\%$ breaking of HADS.   
}
\label{results1}
\begin{tabular}{c c c c c c c c}
  \hline \hline
     Molecule   &~~~~ $J^P$&  Threshold   &~~~~B.E.(Scenario A)   &~~~~B.E.(Scenario B) 
         \\ \hline   $\bar{D}T_{\bar{c}\bar{c}}^{0}$   &~~~~ $0^-$ & 5900.3 &~~~~   $24.1_{-22.5}^{+39.4}(16.5_{-12.1}^{+16.1})$  &~~~  $ 32.1_{-28.3}^{+44.9}(23.3_{-15.4}^{+19.3}) $
         \\  $\bar{D}T_{\bar{c}\bar{c}}^{1}$  &~~~~ {$1^-$}&  5978.3 &~~~~$24.7_{-23.0}^{+39.7}(16.8_{-12.2}^{+16.2})$   &~~~   $32.8_{-28.8}^{+45.3}(23.6_{-15.5}^{+19.4})$
         \\  $\bar{D}T_{\bar{c}\bar{c}}^{2}$   &~~~~ $2^-$&  6037.7 &~~~~ $25.2_{-23.3}^{+40.0}(17.0_{-12.3}^{+16.2})$  &~~~   $33.4_{-29.1}^{+45.5}(23.8_{-15.6}^{+19.4}) $ 
         \\   $\bar{D}^{\ast}T_{\bar{c}\bar{c}}^{0}$   &~~~~ $1^-$& 6042.3  &~~~~$29.5_{-26.2}^{+42.1}(18.6_{-13.0}^{+16.7})$   &~~~  $ 38.2_{-32.0}^{+47.6}(25.6_{-16.3}^{+19.9})$
        \\   $\bar{D}^{\ast}T_{\bar{c}\bar{c}}^{1}$  &~~~~ {$0^-$}& 6120.3   &~~~~$43.7_{-35.3}^{+50.7}(29.7_{-18.0}^{+21.6})$ &~~~   $26.1_{-23.8}^{+39.6}(15.6_{-11.4}^{+15.2}) $
         \\   $\bar{D}^{\ast}T_{\bar{c}\bar{c}}^{1}$  &~~~~ {$1^-$}& 6120.3  &~~~~$36.8_{-31.0}^{+46.6}(24.1_{-15.6}^{+19.2})$  &~~~ $ 32.4_{-28.1}^{+43.8}(20.6_{-13.9}^{+17.6})$
         \\$\bar{D}^{\ast}T_{\bar{c}\bar{c}}^{1}$  &~~~~ {$2^-$}&   6120.3  &~~~~$24.2_{-22.3}^{+38.2}(14.0_{-10.6}^{+14.4})$  &~~~   $46.1_{-36.7}^{+52.1}(31.6_{-18.8}^{+22.4}) $
         \\
         $\bar{D}^{\ast}T_{\bar{c}\bar{c}}^{2}$  &~~~~ {$1^-$}&  6179.7  &~~~~ $51.7_{-39.8}^{+55.0}(35.8_{-20.5}^{+24.0})$   &~~~  $20.9_{-19.8}^{+35.7}(11.2_{-9.0}^{+12.8})$
         \\   $\bar{D}^{\ast}T_{\bar{c}\bar{c}}^{2}$  &~~~~ {$2^-$}&  6179.7  &~~~~$37.4_{-31.3}^{+46.8}(24.3_{-15.6}^{+19.3}) $  &~~~ $32.9_{-28.5}^{+44.0}(20.8_{-14.0}^{+17.7})$ 
         \\$\bar{D}^{\ast}T_{\bar{c}\bar{c}}^{2}$  &~~~~ {$3^-$}&    6179.7     &~~~~$19.0_{-18.3}^{+34.3}(9.8_{-8.1}^{+12.0})$   &~~~  $54.3_{-41.3}^{+56.4}(37.8_{-21.3}^{+24.8}) $
         \\
  \hline \hline
\end{tabular}
\end{table}

    First,  we search for poles of $\bar{D}^{(\ast)}T_{\bar{c}\bar{c}}^{(\ast)}$  with  single-channel potentials.  In Table~\ref{results1}, we present the poles found in Scenario A and Scenario B, where the values inside and outside the brackets correspond to the cutoff $\Lambda=0.75$ GeV and $\Lambda=1.5$ GeV, respectively. One can see that our results are weakly dependent on the cutoff. Even taking into account the $25\%$ uncertainty of HADS,  all of these states are below the $\bar{D}^{(\ast)}T_{\bar{c}\bar{c}}^{(\ast)}$ mass thresholds. 
    Obviously, we obtain a complete HQSS multiplet of hadronic molecules composed of a  $\bar{D}^{(\ast)}$ meson and a $T_{\bar{c}\bar{c}}^{(\ast)}$ tetraquark state. In other words, the existence of $\bar{D}^{(\ast)}\Sigma_{c}^{(\ast)}$ hadronic molecules implies the existence of  $\bar{D}^{(\ast)}T_{\bar{c}\bar{c}}^{(\ast)}$ molecules, if HQSS is approximately satisfied.  We note in passing that  the existence of $\bar{D}^{(\ast)}\Sigma_{c}^{(\ast)}$ molecules also implies the existence of a complete HQSS multiplet of  $\Xi_{cc}^{(\ast)}\Sigma_{c}^{(\ast)}$ hadronic molecules in terms of HADS~\cite{Pan:2019skd}, which agrees with the lattice QCD simulations~\cite{Junnarkar:2019equ}. 

\begin{table}[!h]
\centering
\caption{Pole positions relative to the corresponding two-body  thresholds (in units of MeV) , $M-i\Gamma/2$, for the $\bar{D}^{(\ast)}T_{\bar{c}\bar{c}}^{(\ast)}$ molecules  in Scenario A and Scenario B obtained with coupled-channel potentials. The numbers inside and outside brackets correspond to $\Lambda=0.75$ GeV and $\Lambda=1.5$ GeV. The superscripts and subscripts represent corrections induced by a $25\%$ breaking of HADS.   
}
\label{results2}
\footnotesize
\begin{tabular}{ c c c c c c c c}
  \hline \hline
      Molecule   &~ $J^P$ &  Threshold&~$M_\mathrm{  threshold}-M$(Scenario A)   &~ $M_\mathrm{threshold}-M$(Scenario B) 
         \\ \hline   
         $\bar{D}T_{\bar{c}\bar{c}}^{0}$   &~$0^-$&  5900.3  &~  $26.3^{+42.6}_{-24.4} (17.9^{+17.4}_{-13.0})$  &~  $34.2^{+47.4}_{-30.0} (24.6^{+20.4}_{-16.2}) $
         \\  $\bar{D}T_{\bar{c}\bar{c}}^{1}$  &~ {$1^-$}& 5978.3  &~$29.0^{+46.4}_{-26.7} (19.4^{+18.9
         }_{-14.0})$   &~  $36.4^{+49.6}_{-31.6} (25.9^{+21.3}_{-17.0})$
         \\  $\bar{D}T_{\bar{c}\bar{c}}^{2}$   &~$2^-$&  6037.7 &~ $29.2^{+46.0}_{-26.7} (19.4^{+18.7
         }_{-13.9})$  &~ $36.8^{+49.6}_{-31.8} (26.0^{+21.3
         }_{-17.0}) $  
         \\   $\bar{D}^{\ast}T_{\bar{c}\bar{c}}^{0}$   &~ $1^-$ & 6042.3 &~$31.7^{+43.4}_{-28.0} + 1.2^{-1.2}_{-0.8} i$  $(19.7^{+17.8}_{-13.7} + 1.0^{+0.3}_{-0.5} i)$ &~ $ 40.9^{+51.6}_{-34.0} + 0.5^{-0.5}_{-0.2} i(27.2^{+21.4}_{-17.3} + 0.5^{-0.1}_{-0.1} i)$
        \\   $\bar{D}^{\ast}T_{\bar{c}\bar{c}}^{1}$  &~{$0^-$}&  6120.3  &~$42.0^{+48.5}_{-34.0} + 2.0^{+1.0}_{-1.3} i (28.0^{+20.4}_{-17.1} + 0.9^{+0.6}_{-0.5} i )$ &~  $24.8^{+37.9}_{-22.7} + 1.5^{+0.9}_{-1.1} i (14.3^{+14.1}_{-10.6} + 0.6^{+0.4}_{-0.4} i ) $
         \\   $\bar{D}^{\ast}T_{\bar{c}\bar{c}}^{1}$  &~ {$1^-$} & 6120.3 &~$36.5^{+46.4}_{-30.8} + 1.8^{-1.8}_{-0.9} i (23.3^{+19.0}_{-15.1} + 1.4^{+0.3}_{-0.6} i )$  &~ $30.9^{+41.2}_{-26.8} + 2.4^{-2.0}_{-1.5} i (19.0^{+16.2}_{-12.9} + 1.7^{+0.7}_{-0.9} i )$
         \\$\bar{D}^{\ast}T_{\bar{c}\bar{c}}^{1}$  &~{$2^-$}& 6120.3 &~$25.7^{+40.4}_{-23.6} + 1.1^{+0.2}_{-0.9} i (14.7^{+15.1}_{-11.1} + 0.8^{+0.5}_{-0.5} i )$  &~  $48.1^{+53.9}_{-38.1} + 0.3^{-0.3}_{-0.1} i (32.8^{+23.4}_{-19.5} + 0.3^{-0.0}_{-0.1} i ) $
         \\
         $\bar{D}^{\ast}T_{\bar{c}\bar{c}}^{2}$  &~{$1^-$}&  6179.7  &~$49.8^{+49.5}_{-38.2} + 1.6^{+0.3}_{-0.8} i (34.1^{+22.5}_{-19.5} + 1.2^{+0.1}_{-0.5} i )$   &~ $19.8^{+34.4}_{-18.8} + 0.9^{+0.1}_{-0.7} i (10.2^{+12.0}_{-8.3} + 0.5^{+0.3}_{-0.3} i )$
         \\   $\bar{D}^{\ast}T_{\bar{c}\bar{c}}^{2}$  &~ {$2^-$}&  6179.7 &~$25.3^{+41.6}_{-28.9} + 2.4^{-1.0}_{-1.4} i (21.7^{+17.1}_{-14.0} + 1.7^{+0.5}_{-0.8} i )$   &~$ 30.1^{+39.6}_{-26.2} + 2.4^{-0.2}_{-1.5} i (18.3^{+15.6}_{-12.4} + 1.7^{+0.6}_{-0.8} i ) $
         \\$\bar{D}^{\ast}T_{\bar{c}\bar{c}}^{2}$  &~{$3^-$}&  6179.7  &~~~~$19.0^{+34.3}_{-18.3} (9.8^{+12.0}_{-8.1})$   &~~~ $ 54.3^{+56.4}_{-41.3} (37.8^{+24.8}_{-21.3})$ 
         \\
  \hline \hline
\end{tabular}
\end{table}

Next we take  into account the inelastic potentials,  and   search for  poles  with coupled-channel potentials.  We present the results   in Table~\ref{results2}, where some poles are located in the complex plane instead of  in the real axis due to  the contribution of the inelastic channels. We find that the resulting  widths  are rather small,  similar to the  $\bar{D}^{(\ast)}\Sigma_{c}^{(\ast)}$ molecules~\cite{Sakai:2019qph}.  We stress that the  number of     $\bar{D}^{(\ast)}T_{\bar{c}\bar{c}}^{(\ast)}$ molecules remain unchanged after including the inelastic potentials, and the masses of the $\bar{D}^{(\ast)}T_{\bar{c}\bar{c}}^{(\ast)}$ molecules change little compared with the results shown in Table~\ref{results1}. That is to say, even with coupled-channel potentials,  we still obtain a complete HQSS multiplet of   $\bar{D}^{(\ast)}T_{\bar{c}\bar{c}}^{(\ast)}$ hadronic molecules.     One should note that as we reduce the strength of the $\bar{D}^{(\ast)}T_{\bar{c}\bar{c}}^{(\ast)}$ potential , all the poles are still located in the same Riemann sheet. However, if we increase the strength of the potentials, some poles move to different Riemann sheets, leading to two poles located in the same Riemann sheet.

{ 
 At the quark level, the $\bar{D}^{(\ast)}T_{\bar{c}\bar{c}}^{(\ast)}$  molecules can be viewed as triply charmed hexaquark states. We note that in the molecular picture, such triply charmed hexaquark states can be formed by different combinations of hadrons. \footnote{Of course,   triply charmed hexaquark states can also be compact multiquark states.}  In the following,  we  investigate all the possibilities in detail.    In terms of baryon number, the triply charmed hexaquark states can be classified into two groups: one with baryon number $B=0$ and the other with $B=2$.  In the  $B=2$ case, the triply charmed hexaquark states can be formed by  either the $\Xi_{cc}^{(\ast)}\Sigma_{c}^{(\ast)}$ system or  the $\Omega_{ccc}N$ system, where $N$ represents the proton or neutron.  In Ref.~\cite{Pan:2019skd}, utilizing  HADS we have predicted a series of $\Xi_{cc}^{(\ast)}\Sigma_{c}^{(\ast)}$ hadronic molecules from the existence of   $\bar{D}^{(\ast)}\Sigma_{c}^{(\ast)}$ hadronic molecules. These predictions are based on the assumed molecular nature of the three pentaquark states, $P_{c}(4312)$, $P_{c}(4440)$, and $P_{c}(4457)$, discovered by the LHCb Collaboration in 2019. As for the $\Omega_{ccc}N$ system, we find that it can not bind because of   the repulsive Coulomb force and  the  Okubo-Zweig-Iizuka(OZI)  rule. }

 \begin{figure}[!h]
\begin{center}
\includegraphics[width=6.0in]{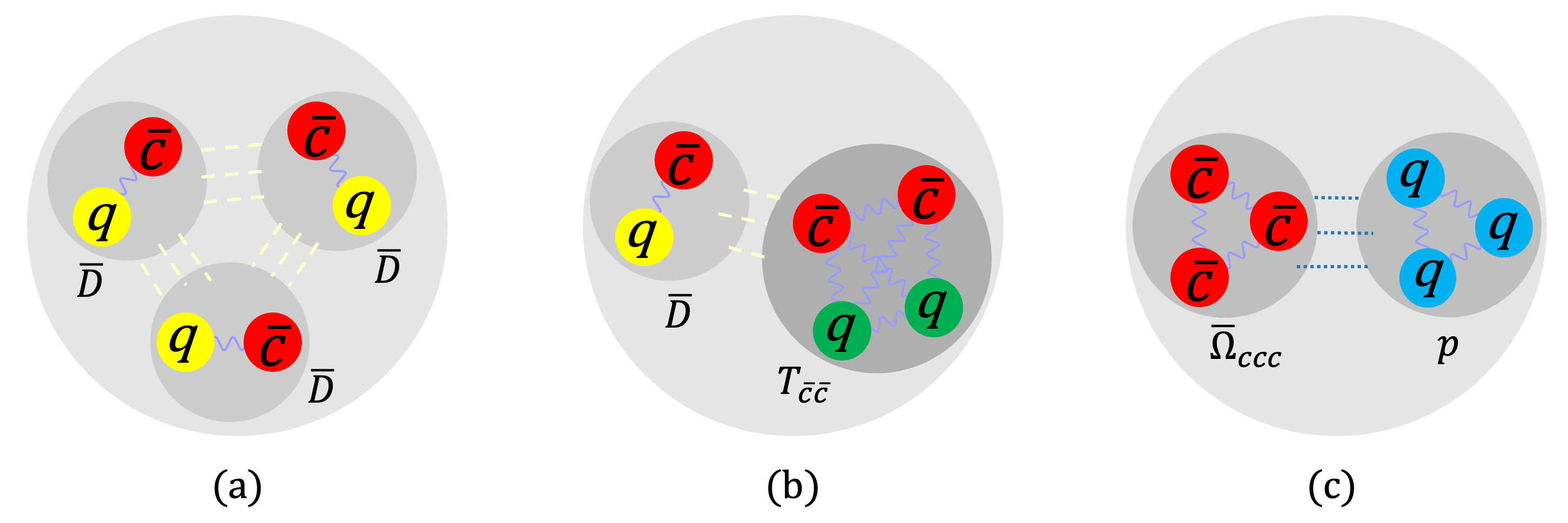}
\caption{ Three possible configurations of triply charmed hexaquark states with the same quark contents:  three charmed mesons $\bar{D}^{(\ast)}\bar{D}^{(\ast)}\bar{D}^{(\ast)}$ (a), a charmed meson and a doubly charmed tetraquark $\bar{D}^{(\ast)}T_{\bar{c}\bar{c}}^{(\ast)}$ (b), and a triply charmed anti-baryon and a proton $\bar{\Omega}_{ccc}{p}$ (c).    }
\label{six}
\end{center}
\end{figure}

 {  
 The  triply charmed hexaquark states with baryon number $B=0$  can exist either in the form of two-body systems  $\bar{D}^{(\ast)}T_{\bar{c}\bar{c}}^{(\ast)}$ and $\bar{\Omega}_{ccc}{p}$,  or the three-body system $\bar{D}^{(\ast)}\bar{D}^{(\ast)}\bar{D}^{(\ast)}$  as shown in Fig.~\ref{six}.  Among the three possible configurations, the existence of  $\bar{D}^{(\ast)}T_{\bar{c}\bar{c}}^{(\ast)}$ molecules have been studied in terms of HADS in the present work, which are based  on  the existence  of $\bar{D}^{(\ast)}\Sigma_{c}^{(\ast)}$ molecules, similar to the existence of  $\Xi_{cc}^{(\ast)}\Sigma_{c}^{(\ast)}$ molecules. In     Refs.~\cite{Wu:2021kbu,Luo:2021ggs}, we have  confirmed  the existence of three-body molecules  $D^{(\ast)}D^{(\ast)}D^{(\ast)}$, which are related to  the molecular nature of $T_{cc}^+(3875)$ recently discovered by the LHCb Collaboration . 
For the $\bar{\Omega}_{ccc}{p}$ system, it is quite difficult to bind by the strong interaction due to the OZI rule, while it can bind by  the Coulomb force as shown below.   }

\begin{figure}[!h]
\begin{center}
\includegraphics[width=4.0in]{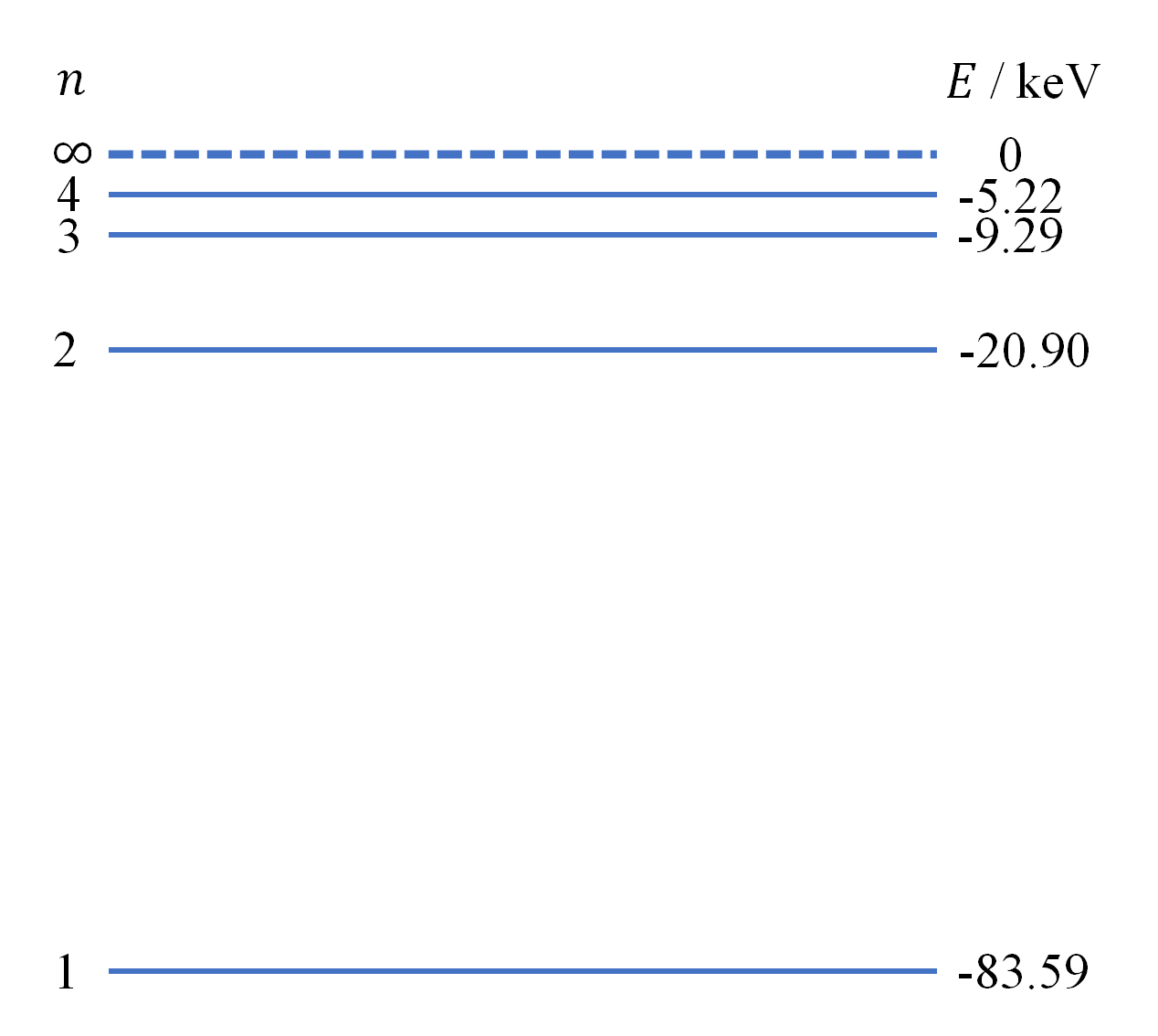}
\caption{ Energy spectrum of the $\bar{\Omega}_{ccc}{p}$ molecule for the orbital angular momentum $l=0$.    }
\label{np}
\end{center}
\end{figure}

The electromagnetic interaction of the $\bar{\Omega}_{ccc}{p}$ system   reads 
\begin{eqnarray}
V_{\bar{\Omega}_{ccc}{p}}^{C}=-2\frac{\alpha}{r},  
\end{eqnarray}
where $\alpha=1/137$ is the fine structure constant. Since the $\Omega_{ccc}$ baryon has not been observed so far,  we take the mass $m_{\Omega_{ccc}}=4795.6$ MeV predicted by lattice QCD~\cite{Lyu:2021qsh}. solving the Schr\"odinger equation via the Gaussian expansion approach, we find that there indeed exists a   shallow $\bar{\Omega}_{ccc}{p}$ bound state with a binding energy of $83.6$ keV  for the orbital angular momentum $l=0$. 
 Recall that for a Hydrogenlike atom, the energy level is given by  
\begin{eqnarray}
E_{n}=-\frac{\mu e^4}{2 \hbar }\frac{Z^2}{n^2},
\end{eqnarray}
 where $\mu$ is the reduced mass of {  the Hydrogenlike atom}, $e$ is the electron charge, and $Z$ is the charge number. Applying the above formula to the $\bar{\Omega}_{ccc}{p}$ system, one obtains the  binding energy $83.6$ keV for the ground state, in agreement with our numerical result.  In Fig.~\ref{np}, we present the energy spectrum  of  the $\bar{\Omega}_{ccc}{p}$ system, which is typical of a two-body system bound by the electromagnetic force. We note that the $\bar{\Omega}_{ccc}{p}$  is similar to the $D^\pm D^{*\mp}$ hadronic atom of Ref.~\cite{Zhang:2020mpi} but differs in that the strong interaction between $\bar{\Omega}_{ccc}$ and ${p}$ is more reduced because of the OZI rule.

{ It should be noted that in the molecular picture the triply charmed  hexaquark states with $B=0$ can have the form of either $\bar{D}^{(\ast)}T_{\bar{c}\bar{c}}^{(\ast)}$, $\bar{D}^{(\ast)}\bar{D}^{(\ast)}\bar{D}^{(\ast)}$,  or $\bar{\Omega}_{ccc}{p}$ as shown in Fig.~\ref{six}. We find that the $\bar{D}^{(\ast)}\bar{D}^{(\ast)}\bar{D}^{(\ast)}$ and  $\bar{D}^{(\ast)}T_{\bar{c}\bar{c}}^{(\ast)}$  molecules are  significantly different from the $\bar{\Omega}_{ccc}{p}$ atom due to their different binding mechanisms. As for the $\bar{D}^{(\ast)}\bar{D}^{(\ast)}\bar{D}^{(\ast)}$ and  $\bar{D}^{(\ast)}T_{\bar{c}\bar{c}}^{(\ast)}$  molecules,  the 
 number of $\bar{D}^{(\ast)}\bar{D}^{(\ast)}\bar{D}^{(\ast)}$ molecules are less than that of  $\bar{D}^{(\ast)}T_{\bar{c}\bar{c}}^{(\ast)}$  molecules due to the constraint of Bose-Einstein statistics. In addition,   the $\bar{D}^{(\ast)}T_{\bar{c}\bar{c}}^{(\ast)}$  molecules can  decay into three charmed mesons~\cite{Eichten:2017ffp}, while the $\bar{D}^{(\ast)}\bar{D}^{(\ast)}\bar{D}^{(\ast)}$ molecules are more likely to decay into three charmed mesons together with a $\pi$ or a $\gamma$~\cite{Wu:2021kbu}, which can be used to discriminate the $\bar{D}^{(\ast)}T_{\bar{c}\bar{c}}^{(\ast)}$  and $\bar{D}^{(\ast)}\bar{D}^{(\ast)}\bar{D}^{(\ast)}$ molecules.      }

\section{Summary and conclusion}
\label{sum}

The pentaquark states,  $P_{c}(4312)$, $P_{c}(4440)$, and $P_{c}(4457)$,  discovered by the LHCb Collaboration, have been suggested to be  $\bar{D}^{(\ast)}\Sigma_{c}$ hadronic molecules. Together with the other four states dictated by HQSS, a complete HQSS multiplet of hadronic molecules may exist. HADS allows one to relate the $\bar{D}^{(\ast)}T_{\bar{c}\bar{c}}^{(\ast)}$ system to the $\bar{D}^{(\ast)}\Sigma_{c}^{(\ast)}$ system.  The $T_{\bar{c}\bar{c}}^{(\ast)}$  are isovector doubly charmed  compact tetraquark states,  which have been extensively studied after the discovery of $T_{cc}^{+}$ by the LHCb Collaboration. Since the masses of the isovector tetraquark states  $T_{\bar{c}\bar{c}}^{(\ast)}$ are  unknown, we took the average masses estimated by several theoretical models, which are the main uncertainty of our present work.  

In terms of HADS, the $\bar{D}^{(\ast)}T_{\bar{c}\bar{c}}^{(\ast)}$ potentials are  the same as the $\bar{D}^{(\ast)}\Sigma_{c}^{(\ast)}$ potentials. We determined the unknown parameters of the $\bar{D}^{(\ast)}T_{\bar{c}\bar{c}}^{(\ast)}$ potentials  by reproducing the masses of $P_{c}(4440)$ and  $P_{c}(4457)$ with two spin assignments. We then employed the Lippmann-Schwinger equation to search for  poles near the mass thresholds of $\bar{D}^{(\ast)}T_{\bar{c}\bar{c}}^{(\ast)}$  with  single-channel  and coupled-channel potentials, respectively. We find that,  even taking into account a $25\%$ breaking of HADS,  there exists a complete HQSS multiplet of $\bar{D}^{(\ast)}T_{\bar{c}\bar{c}}^{(\ast)}$ hadronic molecules  composed of a charmed meson and a doubly charmed compact tetraquark state. Such   states are a new kind of hadronic molecules, which, if discovered, will further enlarge the family of hadronic molecules. { At last, we summarised the triply charmed hexaquark states formed by two kind of combinations of hadrons,  $\bar{D}^{(\ast)}T_{\bar{c}\bar{c}}^{(\ast)}$ two-body hadronic molecules, $\bar{D}^{(\ast)}\bar{D}^{(\ast)}\bar{D}^{(\ast)}$ three-body hadronic molecules, or  $\bar{\Omega}_{ccc}{p}$  hadronic atom with baryon number 0 and  $\Xi_{cc}^{(\ast)}\Sigma_{c}^{(\ast)}$ two-body hadronic molecules with baryon number $B=2$. In particular,   }   we predicted the existence of a different kind of triply charmed hadronic  
atom of $\bar{\Omega}_{ccc}{p}$ held together solely by the  Coulomb force. It is interesting to note that the same mechanism could also lead to the existence of other hadrnoic atoms as well, such as  $\Omega_{ccc}\Omega_{sss}$,  $\Omega_{bbb} p$, and $\Omega_{bbb}\Omega_{ccc}$.

\section{Acknowledgments}
  This work is supported in part by the National Natural Science Foundation of China under Grants No.11975041,  No.11735003, and No.11961141004. Ming-Zhu Liu acknowledges support from the National Natural Science Foundation of
China under Grant No.12105007 and  China Postdoctoral
Science Foundation under Grants No. 2022M710317, and No. 2022T150036.  Tian-Wei Wu acknowledges support from the National Natural Science Foundation of China under Grant No.12147152.

\appendix

\section{ Contact-range potentials for the $T_{\bar{c}\bar{c}}^{(\ast)}\bar{D}^{(\ast)}$ system }

The $T_{\bar{c}\bar{c}}^{(\ast)}\bar{D}^{(\ast)}$ states with spin $J=0,1,2,3$ are expressed in terms of the total spin of the light quark and heavy quark pairs as follows:
\begin{eqnarray}
|T_{\bar{c}\bar{c}}^{0}\bar{D}(0^{-})\rangle &=& \frac{1}{\sqrt{3}}{1/2}_{H}\otimes
{1/2}_{L}+ \sqrt{\frac{2}{3}}{3/2}_{H}\otimes
{3/2}_{L},   \\ \nonumber
|T_{\bar{c}\bar{c}}^{1}\bar{D}(1^{-})\rangle &=&
\frac{\sqrt{2}}{3}{1/2}_{H}\otimes
{1/2}_{L}-\frac{1}{3}{1/2}_{H}\otimes
{3/2}_{L} +\frac{1}{3}{3/2}_{H}\otimes
{1/2}_{L}+\frac{\sqrt{5}}{3}{3/2}_{H}\otimes
{3/2}_{L}, \\ \nonumber
|T_{\bar{c}\bar{c}}^{2}\bar{D}(2^{-})\rangle &=&
\frac{1}{\sqrt{3}}{3/2}_{H}\otimes
{1/2}_{L}-\frac{1}{\sqrt{3}}{1/2}_{H}\otimes
{3/2}_{L}+\frac{1}{\sqrt{3}}{3/2}_{H}\otimes
{3/2}_{L},  \\ \nonumber
|T_{\bar{c}\bar{c}}^{0}\bar{D}^{\ast}(1^{-})\rangle &=&
-\frac{1}{3\sqrt{3}}{1/2}_{H}\otimes
{1/2}_{L} +\frac{2\sqrt{\frac{2}{3}}}{3}{1/2}_{H}\otimes
{3/2}_{L}-\frac{2\sqrt{\frac{2}{3}}}{3}{3/2}_{H}\otimes
{1/2}_{L}+\frac{\sqrt{\frac{10}{3}}}{3}{3/2}_{H}\otimes
{3/2}_{L},  \\ \nonumber
|T_{\bar{c}\bar{c}}^{1}\bar{D}^{\ast}(0^{-})\rangle &=& \sqrt{\frac{2}{3}}{1/2}_{H}\otimes
{1/2}_{L}-\frac{1}{\sqrt{3}}{3/2}_{H}\otimes
{3/2}_{L},   \\ \nonumber
|T_{\bar{c}\bar{c}}^{1}\bar{D}^{\ast}(1^{-})\rangle &=&
\frac{1}{\sqrt{2}}{1/2}_{H}\otimes
{3/2}_{L} +\frac{1}{\sqrt{2}} {3/2}_{H}\otimes
{1/2}_{L}, \\ \nonumber
|T_{\bar{c}\bar{c}}^{1}\bar{D}^{\ast}(2^{-})\rangle &=&
-\frac{1}{\sqrt{6}}{3/2}_{H}\otimes
{1/2}_{L}+\frac{1}{\sqrt{6}}{1/2}_{H}\otimes
{3/2}_{L}+\sqrt{\frac{2}{3}}{3/2}_{H}\otimes
{3/2}_{L},
\\ \nonumber
|T_{\bar{c}\bar{c}}^{2}\bar{D}^{\ast}(1^{-})\rangle &=&
\frac{2\sqrt{\frac{5}{3}}}{3}{1/2}_{H}\otimes
{1/2}_{L}+\frac{\sqrt{\frac{5}{6}}}{3}{1/2}_{H}\otimes
{3/2}_{L} -\frac{\sqrt{\frac{5}{6}}}{3} {3/2}_{H}\otimes
{1/2}_{L}-\frac{\sqrt{\frac{2}{3}}}{3} {3/2}_{H}\otimes
{3/2}_{L},  \\ \nonumber
|T_{\bar{c}\bar{c}}^{2}\bar{D}^{\ast}(2^{-})\rangle &=&
\frac{1}{\sqrt{2}}{1/2}_{H}\otimes
{3/2}_{L} +\frac{1}{\sqrt{2}} {3/2}_{H}\otimes
{1/2}_{L},
\\ \nonumber
|T_{\bar{c}\bar{c}}^{2}\bar{D}^{\ast}(3^{-})\rangle &=&
{3/2}_{H}\otimes
{3/2}_{L}.
\end{eqnarray}

In the heavy quark limit, the interactions are only dependent on the degree of freedom of light quarks. As a result, we can write the  matrix elements of  the potentials as $F_{1/2}=\langle 1/2_{L} | V| 1/2_{L} \rangle$  and $F_{3/2}=\langle 3/2_{L} | V| 3/2_{L} \rangle$, then the   potentials for  the $T_{\bar{c}\bar{c}}^{(\ast)}\bar{D}^{(\ast)}$ system can be expressed in terms of two parameters, e.g., $F_{1/2}$  and $F_{3/2}$.  
For the $T_{\bar{c}\bar{c}}^{0}\bar{D}-T_{\bar{c}\bar{c}}^{1}\bar{D}^{\ast}$ coupled channels, the contact-range potentials $V$ in matrix form read 
\begin{equation}
    V_{T_{\bar{c}\bar{c}}^{0}\bar{D}-T_{\bar{c}\bar{c}}^{1}\bar{D}^{\ast}}^{J=0}=\begin{pmatrix}\frac{1}{3}F_{1/2}+ \frac{2}{3}F_{3/2} &\frac{\sqrt{2}}{3}(F_{1/2}-F_{3/2})
    \\\frac{\sqrt{2}}{3}(F_{1/2}-F_{3/2})& \frac{2}{3}F_{1/2}+\frac{1}{3}F_{3/2} \end{pmatrix}
\end{equation}
For the $T_{\bar{c}\bar{c}}^{1}\bar{D}-T_{\bar{c}\bar{c}}^{0}\bar{D}^{\ast}-T_{\bar{c}\bar{c}}^{1}\bar{D}^{\ast}-T_{\bar{c}\bar{c}}^{2}\bar{D}^{\ast}$ coupled channels, the contact-range potentials $V$ in matrix form read $   V_{T_{\bar{c}\bar{c}}^{1}\bar{D}-T_{\bar{c}\bar{c}}^{0}\bar{D}^{\ast}-T_{\bar{c}\bar{c}}^{1}\bar{D}^{\ast}-T_{\bar{c}\bar{c}}^{2}\bar{D}^{\ast}}^{J=1} =$
\begin{equation}
      \begin{pmatrix}
       \frac{1}{3}F_{1/2}+ \frac{2}{3}F_{3/2}&-\frac{1}{3}\sqrt{\frac{2}{3}}(F_{1/2}-F_{3/2})  & \frac{1}{3\sqrt{2}}(F_{1/2}-F_{3/2})  & \frac{1}{3}\sqrt{\frac{5}{6}}(F_{1/2}-F_{3/2})
    \\-\frac{1}{3}\sqrt{\frac{2}{3}}(F_{1/2}-F_{3/2}) &\frac{1}{3}F_{1/2}+ \frac{2}{3}F_{3/2}  &-\frac{2}{3\sqrt{3}}(F_{1/2}-F_{3/2})  & 0
    \\\frac{1}{3\sqrt{2}}(F_{1/2}-F_{3/2})& -\frac{2}{3\sqrt{3}}(F_{1/2}-F_{3/2})  & \frac{1}{2}F_{1/2}+\frac{1}{2}F_{3/2} & -\frac{1}{6}\sqrt{\frac{5}{3}}(F_{1/2}-F_{3/2}) \\\frac{1}{3}\sqrt{\frac{5}{6}}(F_{1/2}-F_{3/2})&0  & -\frac{1}{6}\sqrt{\frac{5}{3}}(F_{1/2}-F_{3/2}) & \frac{5}{6}F_{1/2}+\frac{1}{6}F_{3/2}
    \end{pmatrix}.
\end{equation}
For the $T_{\bar{c}\bar{c}}^{2}\bar{D}-T_{\bar{c}\bar{c}}^{1}\bar{D}^{\ast}-T_{\bar{c}\bar{c}}^{2}\bar{D}^{\ast}$ coupled channels, the contact-range potentials $V$ in matrix form read 
\begin{equation}
    V_{T_{\bar{c}\bar{c}}^{2}\bar{D}-T_{\bar{c}\bar{c}}^{1}\bar{D}^{\ast}-T_{\bar{c}\bar{c}}^{2}\bar{D}^{\ast}}^{J=2}=\begin{pmatrix}
\frac{1}{3}F_{1/2}+ \frac{2}{3}F_{3/2}  &-\frac{1}{3\sqrt{2}}(F_{1/2}-F_{3/2})  & \frac{1}{\sqrt{6}}(F_{1/2}-F_{3/2})
    \\ -\frac{1}{3\sqrt{2}}(F_{1/2}-F_{3/2})  & \frac{1}{6}F_{1/2}+\frac{5}{6}F_{3/2} &-\frac{1}{2\sqrt{3}}(F_{1/2}-F_{3/2})  \\\frac{1}{\sqrt{6}}(F_{1/2}-F_{3/2}) & -\frac{1}{2\sqrt{3}}(F_{1/2}-F_{3/2})  & \frac{1}{2}F_{1/2}+\frac{1}{2}F_{3/2}
    \end{pmatrix}.
\end{equation}
For the $T_{\bar{c}\bar{c}}^{3}\bar{D}^{\ast}$ single channel, the contact-range potential is $V_{T_{\bar{c}\bar{c}}^{3}\bar{D}^{\ast}}^{J=3}=F_{3/2}$.

\bibliography{omegaccc-dibaryon}

\end{document}